\documentclass[12pt]{article}
\usepackage{amssymb}
\usepackage{cite}
\usepackage{epsf}

\def\action{{\cal S}}
\def\courbure{{\cal R}}

\def\MPlanck{{M_P}}

\def\hmu{{\hat \mu}}
\def\hnu{{\hat \nu}}
\def\hsigma{{\hat \sigma}}

\def\dpar{d_{\scriptscriptstyle\parallel}}
\def\dperp{d_{\scriptscriptstyle\perp}}
\def\Rads{R_{\scriptscriptstyle AdS}}

\def\scpt{\scriptstyle}

\newcommand{\sfrac}[2]{{\textstyle\frac{#1}{#2}}}

\begin{document}

\begin{titlepage}
\begin{flushright}
LBNL 45196\\
UCB-PTH-00/08\\
hep-th/0002130 \\
\end{flushright}

\vskip.5cm
\begin{center}
{\huge{\bf $T$ Self-Dual Transverse Space\\
and \\ \vskip0.4cm
Gravity Trapping}}
\end{center}
\vskip1.5cm

\centerline{\sc C. Grojean }
\vskip 15pt

\centerline{Department of Physics,
University of California, Berkeley, CA 94720}
\vskip 5pt
\centerline{\it and}
\vskip 5pt
\centerline{Theoretical Physics Group,
 Lawrence Berkeley National Laboratory, Berkeley, CA 94720}

\vglue .5truecm

\begin{abstract}
\vskip 3pt

\noindent
We advocate that the orbifold $\mathbb{Z}_2$ symmetry of the gravity trapping
model proposed by Randall and Sundrum can  be seen, in appropriate
coordinates, as a symmetry that exchanges the short distances with the large
ones. Using diffeomorphism invariance, we construct extensions defined by patch
glued together. A singularity occurs at the junction and it is interpreted
as a brane, the jump brane, of codimension one. We give explicit realization in
ten and eleven dimensional supergravity and show that the lower dimensional
Planck scale on the brane is finite. The standard model would be trapped on a
supersymmetric brane located at the origin whereas the jump brane would surround
it at a finite distance. The bulk interactions could transmit the supersymmetry
breaking from the jump brane to the SM brane.

\end{abstract}

\end{titlepage}

\section{Introduction}

Since the works of Kaluza and Klein \cite{KK}, we know that, if
there exists some extra-dimensions
to our universe,
an infinity of massive states will be associated to each usual
4D field. Because these KK modes have not yet been observed, necessarily
their masses must be beyond the experimental range of energies resolved in
accelerators ($\sim 1\, \mbox{TeV}$). That is why the size of extra-dimensions cannot
exceed such a ridiculously tiny scale ($\sim 1\, \mbox{TeV}^{-1}
\sim 10^{-19}\, \mbox{m}$).
However recent progresses in string theories \cite{Strings} have corrected this old
scenario suggesting that the Standard Model gauge interactions are
confined to a four dimensional hypersurface while gravity can still propagate
in the whole {\it bulk} space-time. Since the gravity has not yet been tested
for energy beyond $10^{-4}\, \mbox{eV}$ \cite{Gravity Exp},
the bounds on the size of extra-dimensions
are now much lower ($\sim 1\, \mbox{mm}$). This lack of experimental
data allows for a modification of gravitational interactions at submillimetric
distances {\it i.e.} far away from the Planck scale ($\sim 10^{19}\,$GeV) where
quantum gravity were usually thought to take place. This proposal, in a sense,
nullifies\footnote{More precisely, this proposal translates the gauge hierarchy
problem in energy into its Fourier dual, namely a hierarchy between the size of
extra-dimensions and the electroweak scale.} the gauge hierarchy problem \cite{AADD}.
However this analysis was not yet complete essentially because it assumes
a particular factorizable geometry associated to the higher-dimensional space-time
being a direct product of a 4D space-time with a compact space.
Recently this last assumption has been overcome \cite{RS} unwarping
a very rich potential of physical effects. The most exciting one reveals
the non-incompatibility between non-compact extra-dimensions and experimental
gravity \cite{LR}. The crucial point is the existence, in some curved background,
of a normalizable bound state for the metric fluctuations which can be interpreted as the usual
4D graviton. Of course, there still exists an infinite tower of KK modes, even
a continuum spectrum without gap, but the shape of their wave functions
is such that they almost do not overlap with the 4D graviton and thus maintain
the deviations to the Newton's law in limits which are still very far from
experimental bounds. Subsequent to studies of thin shells in general relativity
\cite{Shell} and their revival in a low-energy $M$-theory context \cite{Mtheory},
a toy model was constructed by Randall and Sundrum that
exhibits the previous properties (see \cite{beforeRS} for previous related works).
The question whether this scenario reproduces
the usual 4D gravity beyond the Newton's law has been addressed in
\cite{RSGravity}. The cosmological aspects have been intensively studied
\cite{RSCosmo}. This model brings new approach to tackle the more
severe hierarchy encountered in physics, namely the cosmological constant problem
\cite{CsteCosmo, CGS3}. Whereas phenomenological aspects of {\it warped} compactifications
with one compact extra-dimension have raised some interesting works \cite{RSpheno},
the case of one and many infinite extra-dimensions still waits for further investigations.

The initial model by Randall--Sundrum involves only one extra-dimension. Several
subsequent works \cite{intersections} extend it by considering many intersecting
codimension one branes. Three papers \cite{TwoExtra} consider branes of codimension two.
In a previous publication \cite{CGS3}, we have proposed an effective action inspired
from the brane construction in supergravity that leads to warped compactification
with many extra-dimensions, however it fails to localize gravity. The present paper
gives a generic method that leads to warped geometry trapping gravity: the idea
consists in taking a solution of the equations of motion in the bulk and using
the diffeomorphism invariance in the transverse space to construct a new solution,
defined by patch, gluing together two slices of the initial solution. By
applying a transformation that exchanges the radial distance to the brane with its
inverse, namely imposing a kind of $T$ duality {\it i.e.} a symmetry between
the short and the long distances, we can keep the region of the space-time
that naturally confines gravity and we throw away the domain where the lower
dimensional Planck mass diverges. The next section is devoted to enlighten
our method while reproducing the RS scenario. In section 3, the Ramond--Ramond
fields of low energy effective action of superstring theories are introduced in the
bulk and our method is used to $T$ dualize the usual $p$-brane solutions of
supergravity.

\section{Randall--Sundrum scenario as a $T$ dualization of the transverse space}

In this section we would like to present our method on a simple
example which leads us to the Randall--Sundrum scenario of gravity trapping.
We will be mainly interested  here in the dynamics of the gravitational
fields assuming that the dynamics of the other fields results in an effective
cosmological constant in the {\it bulk} --- the next section will be devoted
to a more elaborated scenario taking into full account the massless modes
of the low energy effective action of superstring theories.
Thus the space-time physics is governed by the following action\footnote{Our
conventions correspond to a mostly plus Lorentzian signature $(-+\ldots +)$
and the definition of the curvature in terms of the metrics is such that an
Euclidean sphere has a positive curvature.$D$ is the space-time dimension.}:
\begin{equation}
\action_{\mbox{\tiny gravity}}^{\mbox{\tiny bulk}}
= \int d^{D} x\, \sqrt{|g|}\,
\left( \frac{\courbure}{2\kappa^2} - \Lambda_{bk}
\right) \ ,
\end{equation}
It is well known for a long time that, when the cosmological constant
is negative, $\Lambda_{bk}<0$, the solution of Einstein equations corresponds
to an Anti--de--Sitter space-time:
\begin{equation}
	\label{eq:AdS}
ds^2 = \left( \frac{r}{\Rads} \right)^2 \eta_{\mu\nu} \, dx^\mu \otimes dx^\nu
+ \left( \frac{\Rads}{r} \right)^2 dr \otimes dr
\ \ \ \
\mu=0\ldots D-1 .
\end{equation}
where the radius $\Rads$ is related to the bulk cosmological constant by:
\begin{equation}
\Rads^{-2} = - \frac{2 \kappa^2 \Lambda_{bk}}{(D-2)(D-1)}
\ .
\end{equation}

The aim of this section is to describe a method to construct new solutions
to Einstein equations using a regular solution such as the previous one.
These new solutions will develop, on hypersurfaces, some discontinuities
in the first derivatives of the metric which will be interpreted as branes.
In the vain of the works of extra-dimensions, we are looking for solutions
that preserve a Poincar\'e invariance in some space-time directions hereafter
called longitudinal directions and that could be identified as the dimensions
associated to our world; the remaining dimensions will be extra-dimensions
transverse to us. Notice that the $AdS$ solution already exhibits a Poincar\'e
invariance in $D-1$ dimensions. The most general $D$ dimensional metric that
preserves a {\it Poincar\'e}$_{D-1}$ symmetry can be written as:
\begin{equation}
ds^2 = A^2(r)\, \eta_{\mu\nu} \, dx^\mu \otimes dx^\nu
+ B^2(r)\, dr \otimes dr
\ .
\end{equation}
In terms of the two functions $A$ and $B$, the Einstein equations read:
\begin{eqnarray}
\label{eq:Einstein1}
& \displaystyle
(D-2) \,\frac{A^{\prime\prime}}{A} +
\frac{(D-2)(D-3)}{2} \left( \frac{A^\prime}{A} \right)^2
- (D-2)\, \frac{A^\prime}{A} \frac{B^\prime}{B} =
-\kappa^2  \Lambda_{bk} B^2
\ ;\\
\label{eq:Einstein2}
& \displaystyle
\frac{(D-2)(D-1)}{2} \left( \frac{A^\prime}{A} \right)^2
=
- \kappa^2 \Lambda_{bk} \, B^2
\ ,
\end{eqnarray}
where  primes denote derivatives with respect to
the transverse coordinate $r$. The $AdS$ solutions simply corresponds
to $A_{AdS}(r)=r/\Rads$ and $B_{AdS}(r)=\Rads/r$.

The key observation is that, even after requiring a Poincar\'e invariance,
the equations of motion still possess a reparametrization invariance in the
transverse space. In our one extra-dimension example, this invariance is associated
to diffeomorphism in the coordinate $r$ and insures that, if
$A_\circ (r)$ and $B_\circ (r)$ are a solution of the Einstein
equations (\ref{eq:Einstein1})-(\ref{eq:Einstein2}),
thus ${\tilde A} (\tilde r) = A_\circ (f(\tilde r))$
and ${\tilde B} (\tilde r) = \pm B_\circ (f(\tilde r)) f^\prime(\tilde r)$
are also a solution, as it can be explicitly checked, for any diffeomorphism
$f$ whose image falls in the support of the original $A_\circ$ and $B_\circ$.
Of course these two solutions correspond to the same physical space if they are used
for covering the whole space-time. However they can be used separately
in order to construct new solution defined by patch on two non-overlapping regions:
this construction mimics the procedure used by Randall and Sundrum and consists
in taking two identical slices of space-time and gluing them together.
Explicitly, the solution can be defined, starting from any
solution $A_\circ(r)$ and $B_\circ(r)$, as:
\begin{eqnarray}
	\label{eq:patch1}
\mbox{for } & r\leq r_\circ & \ \ \
A(r) = A_\circ (r) \ \
B(r) = B_\circ (r)
\\
	\label{eq:patch2}
\mbox{for } & r\geq r_\circ & \ \ \
A(r) = A_\circ (r_\circ f(r)/f(r_\circ)) \ \
B(r) = \pm B_\circ (r_\circ f(r)/f(r_\circ)) \frac{r_\circ f^\prime (r)}{f(r_\circ)}
\end{eqnarray}
The requirement  that the metric remains continuous at $r_\circ$ gives
a constraint between $r_\circ$ and the function $f$, namely:
\begin{equation}
	\label{eq:continuity}
 \frac{r_\circ f^\prime (r_\circ)}{f(r_\circ)} = \pm 1 \ .
\end{equation}
In the present case where $r$ represents a radial distance that remains
positive, there is a particular change of coordinate that fulfills the constraint
(\ref{eq:continuity}) whatever the value of $r_\circ$ is: $f(r) = r_\circ^2/r$.
The significance of this change of coordinate is very clear: the original solution
near infinity is cut and replaced by a copy of the origin and the long distance
solution becomes just a mirror of the short distance region. In that sense
the new solution is {\it $T$ self-dual}. Notice that
the $r \leftrightarrow r_\circ^2/r$ symmetry
is just a $\mathbb{Z}_2$ symmetry in the Randall--Sundrum coordinate,
$y=- R \ln (r/\Rads)$, when $r_\circ=\Rads$. In the RS coordinate, the full
$AdS$ metric (\ref{eq:AdS}) reads:
\begin{equation}
	\label{RSAdS}
ds^2 = e^{-2 y/\Rads} \eta_{\mu\nu} \, dx^\mu \otimes dx^\nu
+ dy \otimes dy .
\end{equation}
Randall and Sundrum have looked for a $\mathbb{Z}_2$ symmetric configuration and have
obtained:
\begin{equation}
ds^2 = e^{-2 |y|/\Rads} \eta_{\mu\nu} \, dx^\mu \otimes dx^\nu
+ dy \otimes dy .
\end{equation}
when a brane with a positive and fine-tuned cosmological constant is placed at
$y=0$ {\it i.e.} $r=\Rads$. This  $\mathbb{Z}_2$ symmetrization is nothing
but the procedure of $T$ dualization of the transverse space described above:
the region of negative $y$, that would correspond to $r\geq \Rads$, is cut
and replaced by a copy of the region of positive $y$ {\it i.e.}
$r\leq \Rads$. This procedure of cutting and pasting is not specific to
the $T$ dualization of transverse space, for instance it has been used
in the first reference in \cite{intersections} to generalize the RS construction
with cosmological constants on a setup of intersecting codimension one
branes. The notion of $T$ transformation will be important for trapping gravity
on higher codimension branes like those appearing in supergravity theories.

The diffeomorphism invariance insures that (\ref{eq:patch1})-(\ref{eq:patch2})
is a solution of the Einstein equations in the bulk. However, even if the
metric is continuous at $r=r_\circ$, its first derivatives are usually not and thus
a Dirac singularity appears in the left hand side of (\ref{eq:Einstein1})
which has to be associated to a singular stress-energy tensor. In our example
it is not difficult to see that this singular stress-energy tensor can be derived
from a term interpreted as a cosmological constant on the hypersurface $r=r_\circ$:
\begin{equation}
	\label{eq:Lambdabr}
\action^{\mbox{\tiny brane}}_{eff}
= - \int d^{D} x\, \sqrt{|g|}\,
 \Lambda_{br}\
\delta \left(
\sqrt{|g_{rr}|}(r-r_\circ) \right)
\end{equation}
where $\Lambda_{br}$ is given by:
\begin{equation}
\Lambda_{br}
=\sqrt{-\frac{8(D-2)}{(D-1)\kappa^2} \Lambda_{bk}}
\ .
\end{equation}

The above procedure of $T$ dualization has the nice property to lead
to a finite $D-1$ dimensional Planck mass. Indeed, whereas in the original
solution it would be:
\begin{equation}
\MPlanck^{D-3} =
\frac{1}{\kappa^2} \int_{0}^{\infty}
dr \, A^{D-3} (r) \, B(r)=
\int_{0}^{\infty}
dr \, \left( \frac{r}{\Rads} \right)^{D-4}
\end{equation}
which diverges near infinity, our solution that throws away the region near
infinity and paste a copy of the region near the origin, naturally gives
a finite   $D-1$ dimensional Planck mass\footnote{Notice that we could have
cut the horizon of $AdS$ and kept two copies of the infinite boundary but
this geometry would not lead to a finite lower dimensional Planck scale.}:
\begin{equation}
	\label{eq:MPl}
\MPlanck^{D-3} =
\frac{1}{\kappa^2}
\left(
\int_{0}^{r_\circ}
dr \,  \left( \frac{r}{\Rads} \right)^{D-4}
+
\int_{r_\circ}^{\infty}
dr \,  \frac{r_\circ^{2D-6}}{r^{D-2}\Rads^{D-4}}
\right)
=
\frac{2}{(D-3)\kappa^2}\, r_\circ \left( \frac{r_\circ}{\Rads} \right)^{D-4}
\end{equation}
At this stage, the position $r_\circ$ of the fixed point under the $T$ symmetry is
arbitrary. It is believed that a dynamical description of the brane beyond its
effective description in terms of a cosmological constant (\ref{eq:Lambdabr}) should
allow to stabilize the value of $r_\circ$. Notice that, when $r_\circ=\Rads$,
the expression (\ref{eq:MPl}) coincides with the Planck mass on the brane computed in
the RS model.

In the next section, we extend our procedure of $\ T$ dualization to solutions
of the equations of motion of supergravity.

\section{Gravity trapping from the branes of supergravity}

In this section, we apply our procedure of $T$ dualization to the
brane solutions of supergravity theories. An electric $p$-brane couples
to a ($p+1$) differential form. While preserving a Poincar\'e invariance
in the dimensions parallel to the brane, the electric field strength curves
the transverse space. Nice solutions of the equations of motion have been
constructed \cite{SugraBranes}. Their remarkable supersymmetric and BPS properties
insure their stability. They are interpreted as collective excitations
of perturbative string theories and they become the elementary objects of dual
theories capturing part of the non-perturbative aspects.
In \cite{CGS3}, it was argued that the brane configurations can be seen
as warped geometry of space-time. Unfortunately, the shape of the warp factor
along the infinite extra-dimension associated to the radial distance to the brane
in the transverse space does not localize gravity as in the RS scenario.
The origin of such a discrepancy is due to the geometry of space-time far away
from the brane. For example in the particular case of vanishing dilaton coupling,
the geometry corresponds\footnote{Usually the solution constructed in supergravity
theories asymptotes $AdS_{\dpar+1}\times S^{\dperp-1}$ near the brane only and it
is normalized such as to recover a $D$ dimensional Minkowskian flat space
at infinity. As we will see in eq. (\ref{eq:H}), this normalization corresponds
to a particular choice of constant of integration.
$AdS_{\dpar+1}\times S^{\dperp-1}$ provides also a solution in the full space.
The physical relevance of this solution is suggested by the fact that the
dynamics of a brane becomes free near the conformal boundary of $AdS$
\cite{GM}. Our argument concerning the gravity localization is unchanged
if the solution with a flat space at infinity is considered since the $\dpar$ Planck
mass also diverges in that case.}
to a product $AdS_{\dpar+1}\times S^{\dperp-1}$, where $\dpar$ is
the dimension of the longitudinal space and $\dperp$, the dimension of the transverse
space, $\dpar+\dperp=D$. Whereas the region near the brane is the horizon of $AdS$,
the region near infinity is associated to the conformal boundary of $AdS$ which
is precisely the part of space-time cut in the RS configuration.
Our solution will consist in $T$ dualizing the $AdS$ horizon in the region near
infinity. A new singularity will appear where the two slices are glued
and we will describe this singularity.

We begin with a review of the brane construction in supergravity theories.
A $p$-brane is coupled to the low-energy effective theory of superstrings.
Below the fundamental energy scale, identified as the energy of the first massive
excitations of the string, the theory can be described by supergravity theories
whose bosonic spectrum contains the metric, a scalar field (the dilaton)
and numerous differential forms.
The bosonic effective action, in supergravity units where the curvature term
is canonically normalized, takes the general form
(we will use $\kappa^2=M^{2-D}$ where $M$ is the Planck mass in ten or eleven
dimensions):
\begin{equation}
	\label{eq:sugra}
\action_{eff}^{\mbox{\tiny sugra}} = \int d^Dx \, \sqrt{|g|}
\left(
\frac{1}{2\kappa^2}\, \courbure 
-\frac{1}{2}\, \partial_\hmu \Phi \partial^\hmu \Phi
- \sum_{n}
\frac{1}{(n+2)!}\, e^{\alpha_{n} \Phi}
F_{\hsigma_1\ldots \hsigma_{n+2}}  F^{\hsigma_1\ldots \hsigma_{n+2}}
\right)
\ ,
\end{equation}
where $ F_{\hmu_1\ldots\hmu_{n+2}} = (n+2) \,
	\partial_{[\hmu_1}\, C_{\hmu_2\ldots \hmu_{n+2}]} $ is the
field strength of the ($n+1$)-differential form $C$, whose coupling
to the dilaton is measured by the coefficient $\alpha_n$.
The allowed values of $n$ depends on the theory we consider.
The coefficients $\alpha_n$ are explicitly
determined by a string computation:
the coupling of the dilaton to differential forms
from the Ramond-Ramond sector
appears at one loop and thus
$\alpha_n^{RR}/\sqrt{2\kappa^2}=(3-n)/2$ in supergravity units,
while the Neveu-Schwarz--Neveu-Schwarz two-form couples at tree level, so
$\alpha_{1}^{NS}/\sqrt{2\kappa^2}=-1$.
In some cases, we can also add a Chern--Simons term
($C\wedge F \wedge F$) to the action, but
it does not have any effect on the classical solutions and we will
neglect it in our analysis.

This {\it bulk} effective action can couple to some {\it branes}. And the total
action is:
\begin{equation}
 \action =
\action_{eff}^{\mbox{\tiny sugra}}
+
\action_{eff}^{\mbox{\tiny branes}}
\end{equation}
The equations of motion are derived form this action and can be read
($\hmu,\hnu=0\ldots D-1$):
\begin{eqnarray}
\nonumber
&&
G_{\hmu\hnu} = 
\kappa^2\, \partial_\hmu \Phi \partial_\hnu \Phi
+ \sum_{n}
\frac{2\kappa^2}{(n+1)!} \, e^{\alpha_n \Phi}\,
	F_{\hmu\hsigma_1\ldots\hsigma_{n+1}}
	F_{\hnu}{}^{\hsigma_1\ldots\hsigma_{n+1}}
\\
&&	\label{eq:braneRicci}  \rule{1.5cm}{0pt} 
+\frac{1}{2} \left( 
-\kappa^2 \, \partial_\hsigma \Phi \partial^\hsigma \Phi
- \sum_{n}
\frac{2 \kappa^2 }{(n+2)!} \, e^{\alpha_n \Phi}\,
     F_{\hsigma_1\ldots\hsigma_{n+2}}  F^{\hsigma_1\ldots\hsigma_{n+2}}
\right)\, g_{\hmu\hnu}\,
+ T_{\hmu\hnu}^{br}
\ ;
\\
	 \label{eq:branedilaton} 
& &
D_\hmu D^\hmu \Phi = \sum_{n}
\frac{\alpha_n }{(n+2)!}\, e^{\alpha_n \Phi}\,
F_{\hsigma_1\ldots\hsigma_{n+2}} F^{\hsigma_1\ldots\hsigma_{n+2}}
+ T_\Phi^{br}
\ ;
\\
	\label{eq:braneTenseur}
& &
\partial_{\hmu_0} \left( \sqrt{|g|} \, e^{\alpha_n \Phi}\,
F^{\hmu_0\ldots\hmu_{n+1}}
\right) = J^{\hmu_1\ldots\hmu_{n+1}}_{br}
\ .
\end{eqnarray}
The {\it brane} stress-energy tensor $T_{\hmu\hnu}^{br}$, the electric currents
$J_{br}$
and the dilatonic current $T_\Phi^{br}$ are formally given by:
\begin{eqnarray}
T_{\hmu\hnu}^{br} =  -
\frac{2 \kappa^2}{\sqrt{|g|}}
\frac{\delta \action_{eff}^{\mbox{\tiny branes}}}{\delta g^{\hmu\hnu}}
\ ;
\ \ \
J^{\hmu_1\ldots\hmu_{n+1}}_{br}
=
-\frac{(n+1)!}{2} \frac{\delta \action_{eff}^{\mbox{\tiny branes}}}{
\delta A_{\hmu_1\ldots \hmu_{n+1}}}
\ ;
\ \ \
T_\Phi^{br}
=
- \frac{\delta \action_{eff}^{\mbox{\tiny branes}}}{\delta \Phi}
\ ,
\end{eqnarray}
and can be derived whenever the effective action describing the dynamics of the
branes is known.

We would like now to construct a solution of these equations of motion
with particular symmetries namely
a Poincar\'e invariance in $\dpar=p+1$ dimensions that will be identified as longitudinal
dimensions and also a rotational invariance in the $\dperp$ dimensional
transverse space. Such a solution has been known for a long time in supergravity
theories and it is expressed as ($\mu,\nu=0\ldots \dpar-1$ and $i,j=1\ldots \dperp$):
\begin{eqnarray}
&&
ds^2  =
H^{2 n_x} \, \eta_{\mu\nu} \, dx^\mu \otimes dx^\nu
+ H^{2 n_y} \,\eta_{ij} \, dy^i \otimes dy^j  \ ;
\\
&&
e^{\Phi}  =
H^{n_\Phi} \ e^{\phi_\circ} \ \ \
(\phi_\circ \ \mbox{ is a constant}) \ ;
\\
&&
C_{\mu_1\ldots\mu_{p+1}} =
- \epsilon_{\mu_1 \ldots \mu_{p+1}}
\frac{1}{{\cal A}_{WZ}}\,
e^{-\alpha_p \phi_\infty /2}\,  H^{-1} \ ;
\end{eqnarray}
where $H$ is a function of the radial distance in the transverse space only.
Notice that the {\it Poincar\'e}$_{\dpar}$ symmetry allows only for a non-vanishing
$\dpar$ differential form and furthermore all the fermionic fields have to vanish.
The consistency of the whole set of equations of motion determines the powers
$n_x$, $n_y$ and $n_\Phi$:
\begin{equation}
	\label{eq:powers}
n_x = -\frac{2 (\dperp-2)\kappa^2}{(\dpar+\dperp-2)\, {\cal A}_{WZ}^2}
\ \ \ \
n_y =   \frac{2 \dpar \kappa^2}{(\dpar+\dperp-2)\, {\cal A}_{WZ}^2}
\ \ \ \
n_\Phi = \frac{\alpha_p}{{\cal A}_{WZ}^2} \ ,
\end{equation}
and the coefficient ${\cal A}_{WZ}$ which has to be related to the dilaton
coupling by:
\begin{equation}
	\label{eq:WZ}
{\cal A}_{WZ}^2 = 2 \kappa^2 \frac{\dpar(\dperp-2)}{\dpar+\dperp-2}
+ \frac{\alpha_p^2}{2} \ .
\end{equation}
In supergravity theories, according to the particular values of the dilaton
coupling previously given, the coefficient ${\cal A}_{WZ}$ is a constant
independent of the dimension of the brane:
\begin{equation}
{\cal A}_{WZ}^2 = 4 \kappa^2 \ .
\end{equation}
The function $H$ is harmonic in the transverse space:
\begin{equation}
	\label{eq:H}
H =  l +  \frac{Q}{r^{\dperp-2}}
\end{equation}
where $l$ is an arbitrary dimensionless constant and $Q$ is a constant with
a dimension $\dperp-2$ in length.

This solution solves the bulk equations of motion everywhere except at the origin
where occurs a singularity, interpreted as a $p$-{\it brane}.
We know exactly the brane action generating such a singularity:
\begin{eqnarray}
	\nonumber
\action_{eff}^{\mbox{\tiny brane}} = T_{p+1} \int d^{p+1} \xi\,
\left(
-\frac{1}{2}
\sqrt{|\gamma|}
\gamma^{ab}
\partial_a X^\hmu \partial_b X^\hnu
g_{\hmu\hnu} (X) \, e^{\beta_p \Phi}
+\frac{p-1}{2} \, \sqrt{|\gamma|}
\right. \\
	\label{eq:actionbraneunivers}
\left.
+\frac{{\cal A}_{WZ}}{(p+1)!}\, \epsilon^{a_1\ldots a_{p+1}}
\, \partial_{a_1} X^{\hmu_1} \ldots \partial_{a_{p+1}} X^{\hmu_{p+1}}
C_{\hmu_1\ldots \hmu_{p+1}}
\right)
\ .
\end{eqnarray}
And the corresponding constant $Q$ in the expression (\ref{eq:H}) of
the harmonic function $H$ is related to the tension $T_{p+1}$ of the brane
by:
\begin{equation}
Q =  \frac{{\cal A}_{WZ}^2 T_{p+1}}{2(\dperp-2)\Omega_{d_\perp-1}} \
e^{-\alpha_p \phi_\circ /2}
\end{equation}
where $\Omega_{\dperp-1}$ is the volume of $S^{\dperp-1}$, the sphere
with $\dperp-1$ angles.

Using this solution, we will now construct a new solution by patch.
The most general solution that respects {\it Poincar\'e}$_{\dpar}\times SO(\dperp)$
can be written as:
\begin{eqnarray}
&&
ds^2  =
A^2 (r) \, \eta_{\mu\nu}\, dx^\mu \otimes dx^\nu
+ B^2 (r) \, dr \otimes dr
+ D^2 (r) d^2 \Omega_{\dperp-1}   \ ;
\\
&&
\Phi  = \Phi (r) ;
\\
&&
C_{\mu_1\ldots\mu_{p+1}} =
-\epsilon_{\mu_1 \ldots \mu_{p+1}}
{\cal A}_{WZ}^{-1}\,
C (r)\ ;
\end{eqnarray}
where $d^2 \Omega_{\dperp-1}={\tilde g}_{\alpha\beta}
d\theta^\alpha \otimes d\theta^\beta$
is the metric on the $S^{\dperp-1}$ described by the angles $\theta^\alpha$,
$\alpha=1\ldots \dperp-1$.
In terms of these functions, the equations of motion
(\ref{eq:braneRicci})-(\ref{eq:braneTenseur}) in the bulk, {\it i.e.} dropping
any singularity, read:
\begin{eqnarray}
\nonumber
&&
\hspace{-.5cm}
\sfrac{(\dpar-1)(\dpar-2)}{2} \left( \frac{A^\prime}{A} \right)^2
+{\scpt (\dpar-1)} \frac{A^{\prime\prime}}{A}
-{\scpt (\dpar-1)} \frac{A^\prime}{A} \frac{B^\prime}{B}
+{\scpt (\dpar-1)(\dperp-1)}  \frac{A^\prime}{A} \frac{D^\prime}{D}
\\
\nonumber
&&
\ \ \
\sfrac{(\dperp-1)(\dperp-2)}{2} \left( \frac{D^\prime}{D} \right)^2
+{\scpt (\dperp-1)} \frac{D^{\prime\prime}}{D}
-{\scpt (\dperp-1)} \frac{B^\prime}{B} \frac{D^\prime}{D}
-\sfrac{(\dperp-1)(\dperp-2)}{2}  \frac{B^2}{D^2}
\\
	\label{eq:Gmunu}
&&
\ \ \  =
-\kappa^2 (\Phi^{\prime})^2
-\kappa^2 {\cal A}_{WZ}^{-2} \, A^{-2 \dpar} \, (C^\prime)^2
\\
\nonumber
&&
\hspace{-.5cm}
\sfrac{\dpar(\dpar-1)}{2} \left( \frac{A^\prime}{A} \right)^2
-{\scpt \dpar(\dperp-1)} \frac{A^\prime}{A} \frac{B^\prime}{B}
-\sfrac{(\dperp-1)(\dperp-2)}{2} \left( \frac{D^\prime}{D} \right)^2
-\sfrac{(\dperp-1)(\dperp-2)}{2}  \frac{B^2}{D^2}
\\
	\label{eq:Grr}
&&
\ \ \ =
\kappa^2 (\Phi^{\prime})^2
-\kappa^2 {\cal A}_{WZ}^{-2} \, A^{-2 \dpar} \, (C^\prime)^2
\\
\nonumber
&&
\hspace{-.5cm}
\sfrac{\dpar(\dpar-1)}{2} \left( \frac{A^\prime}{A} \right)^2
+{\scpt \dpar} \frac{A^{\prime\prime}}{A}
-{\scpt \dpar} \frac{A^\prime}{A} \frac{B^\prime}{B}
+{\scpt \dpar(\dperp-2)}  \frac{A^\prime}{A} \frac{D^\prime}{D}
\\
\nonumber
&&
\ \ \
\sfrac{(\dperp-2)(\dperp-3)}{2} \left( \frac{D^\prime}{D} \right)^2
+{\scpt (\dperp-2)} \frac{D^{\prime\prime}}{D}
-{\scpt (\dperp-2)} \frac{B^\prime}{B} \frac{D^\prime}{D}
-\sfrac{(\dperp-2)(\dperp-3)}{2}  \frac{B^2}{D^2}
\\
	\label{eq:Gab}
&&
\ \ \ =
-\kappa^2 (\Phi^{\prime})^2
+\kappa^2 {\cal A}_{WZ}^{-2} \, A^{-2 \dpar} \, (C^\prime)^2
\\
	 \label{eq:dilaton}
& &
\hspace{-.5cm}
\left( A^{\dpar} B^{-1} D^{\dperp-1} \Phi^\prime \right)^\prime
=
-\alpha_p \, {\cal A}_{WZ}^{-2}  A^{\dpar} B^{-1} D^{\dperp-1} (C^\prime)^2
\\
	\label{eq:tenseur}
& &
\hspace{-.5cm}
\left( A^{-\dpar} B^{-1} D^{\dperp-1} C^\prime \right)^{\prime}
= 0
\end{eqnarray}
where the primes denote derivative with respect to the radial
distance $r$.

The brane solution of supergravity corresponds to
$A_{sg}=H^{n_x}$, $B_{sg}=H^{n_y}$, $D_{sg}=r\, H^{n_y}$,
$e^{\Phi_{sg}}=H^{n_\Phi}\, e^{\phi_\circ}$
and $C_{sg}=H^{-1}\, e^{\phi_\circ/2}$ where $\phi_\circ$ is an arbitrary
constant and $H$ is given in (\ref{eq:H}).
The diffeomorphism invariance insures that
\begin{eqnarray}
\mbox{for } & r\leq r_\circ & \ \ \
\begin{array}{l}
A(r) = H^{n_x} (r) \ \
B(r) = H^{n_y} (r) \ \
D (r) = r\, H^{n_y} (r)
\\
\Phi (r) = \phi_\circ + n_\Phi \ln H (r) \ \
C (r) = e^{\phi_\circ/2}\, H^{-1} (r)
\end{array}
\\
\mbox{for } & r\geq r_\circ & \ \ \
\begin{array}{l}
A(r) = H^{n_x} (r_\circ^2/r) \ \
B(r) = r_\circ^2/r^2 H^{n_y} (r_\circ^2/r)  \ \
D (r) = r_\circ^2/r\, H^{n_y} (r_\circ^2/r)
\\
\Phi (r) = \phi_\circ + n_\Phi \ln H (r_\circ^2/r) \ \
C (r) = e^{\phi_\circ/2}\, H^{-1} (r_\circ^2/r)
\end{array}
\end{eqnarray}
is also a solution of the equations of motion in the bulk, as it can be checked
explicitly. Whereas this $T$ dualization of the transverse space provides a
continuous junction between the two patches, the first derivatives of the fields
have a jump and thus lead to a Dirac singularity. We can interpret this singularity
as a brane of codimension one located at the junction region: we will call it the
{\it jump brane}\footnote{This geometry reminds some aspects of the model
of concentric branes constructed in the third reference in \cite{TwoExtra} with
a discontinuous cosmological constant in a 6D bulk.}. Among the dimensions
on this brane, $\dpar$ ones are non-compact and are parallel to the dimensions
of the brane at the origin, while the $\dperp-1$ remaining directions
are compact and describe a sphere of radius $D(r_\circ)$ and thus at energies
below $D^{-1}(r_\circ)$, the second brane will also appear as a ($\dpar-1$)-brane.
From the explicit expression of the equations of motion, we can derive the
singularity associated to the second brane in terms of the original supergravity
solution:
\begin{eqnarray}
	\label{eq:2ndTmunu}
T_{\mu\nu}^{br} & = &
-2\left( (\dpar-1) \frac{A^\prime_{sg}}{A_{sg}}
+(\dperp-1) \frac{D^\prime_{sg}}{D_{sg}} \right)
\frac{A^2_{sg}}{B^2_{sg}} \eta_{\mu\nu} \, \delta (r-r_\circ)
\\
	\label{eq:2ndTab}
T_{\alpha\beta}^{br} & = &
-2\left( \dpar \frac{A^\prime_{sg}}{A_{sg}}
+(\dperp-2) \frac{D^\prime_{sg}}{D_{sg}} \right)
\frac{D^2_{sg}}{B^2_{sg}} {\tilde g}_{\alpha\beta} \, \delta (r-r_\circ)
\\
	\label{eq:2ndTphi}
T_{\Phi}^{br} & = &
-2 \Phi_{sg}^\prime B_{sg}^{-2} \, \delta (r-r_\circ)
\\
	\label{eq:2ndJ}
J^{\mu_1\ldots\mu_{p+1}}_{br} & = &
2 \epsilon^{\mu_1\ldots\mu_{p+1}} A_{WZ}^{-1} A_{sg}^{-\dpar}
B_{sg}^{-1} D_{sg}^{\dperp-1} C_{sg}^\prime  \, \delta (r-r_\circ)
\end{eqnarray}
where we remind that ${\tilde g}_{\alpha\beta}$ has been defined as the metric
on $S^{\dperp-1}$. An interesting and opening problem that we will not address
in this letter is to determine the effective action $\action_{eff}^{\mbox{\tiny brane}}$
describing the second brane that would lead to the currents
(\ref{eq:2ndTmunu})-(\ref{eq:2ndJ}). In the particular case where the integration
constant, $l$, of the supergravity solution (\ref{eq:H}) has been chosen to vanish,
the stress-energy tensor on the brane can be parametrized by two cosmological
constants along the compact and non-compact directions:
\begin{equation}
T_{\mu\nu}^{br} = -\kappa^2 \Lambda^{br}_{\scriptscriptstyle \parallel}
\, g_{\mu\nu}\, \delta (\sqrt{g_{rr}}(r-r_\circ))
\ \ \
T_{\alpha\beta}^{br} = -\kappa^2 \Lambda^{br}_{\scriptscriptstyle \perp}
\, g_{\mu\nu}\, \delta (\sqrt{g_{rr}}(r-r_\circ))
\end{equation}
where
\begin{eqnarray}
&&
\Lambda^{br}_{\scriptscriptstyle \parallel} \propto
\left( \dperp-1 - 2 (\dperp-2) {\cal A}_{WZ}^{-2}\, \kappa^2 \right)
r_\circ^{-\alpha_p^2 {\cal A}_{WZ}^{-2}/2}
\\
&&
\Lambda^{br}_{\scriptscriptstyle \perp} \propto
(\dperp-2)
r_\circ^{-\alpha_p^2 {\cal A}_{WZ}^{-2}/2}
\end{eqnarray}
Notice that the cosmological constants in the two directions are equal only when
the Wess--Zumino coupling is given by:
\begin{equation}
{\cal A}_{WZ}^{2} = 2(\dperp-2)\kappa^2
\ \ \ \
\mbox{{\it i.e.}}
\ \ \ \
\alpha_p^2 = 4 \kappa^2 \frac{(\dperp-2)^2}{D-2}
\ .
\end{equation}
This is never the case in supergravity.

The most attractive feature of the solution we have just constructed is that, just
as in the RS model, it provides a finite $\dpar$ dimensional Planck mass in spite
of the infinite extra-dimension. Indeed, this scale is now given by:
\begin{equation}
\MPlanck^{\dpar-2}=
\kappa^{-2} \int d^{\dperp} y \, A^{\dpar-2} B D^{\dperp-1}
= 2 \kappa^{-2} \Omega_{\dperp-1} \int_{0}^{r_\circ}
dr \, r^{\dperp-1} H^{4\kappa^2 {\cal A}_{WZ}^{-2}}
=  \kappa^{-2} \Omega_{\dperp-1} Q r_\circ^2
\end{equation}
In the last equality, we have used the supersymmetric value of
${\cal A}_{WZ}$ and set the constant of integration, $l$, to zero.
The fact that $\MPlanck$ converges is a good indication that the gravitational
force will mainly follow a Newton's law in $\dpar$ dimensions, up to deviations
beyond experimental bounds, and suggests the existence of a normalizable bound state
for the metric fluctuations that will be interpreted as a 4D graviton
\cite{CsakiGraviton,FluctGraviton}.
The deviations to the Newton's law can be obtained form the KK spectrum
of the 4D graviton. The equations of motion for the fluctuations will be
greatly simplified by noticing that, in the RS gauge, the stress-energy tensor in
the bulk derived from (\ref{eq:sugra}) satisfies, at the first order in perturbation:
\begin{equation}
T_{\mu\nu}^{(1)} =
\left( T_{\mu\sigma}^{(0)}h_{\rho\nu} + T_{\sigma\nu}^{(0)}h_{\rho\mu} \right)
\eta^{\rho\sigma}
\end{equation}
where
\begin{equation}
ds^2 = A^2 (r) (\eta_{\mu\nu}+h_{\mu\nu}(x,r,\theta)) dx^\mu \otimes dx^\nu
+ B^2 (r) \, dr \otimes dr
+ D^2 (r) d^2 \Omega_{\dperp-1}
\end{equation}
As noticed in \cite{CsakiGraviton}, this relation, that is here rather non-trivial since the stress-energy tensor
is non-linear in the metric, is what is needed to cancel all the non-derivative terms
in $h$ in the Einstein equations.
However, without knowing the effective action for the second brane, it is impossible
to derive the full equations of motion for the fluctuations near the second brane.
We leave this question for further investigations.

Since the supersymmetric extension of the RS model has been debated with
rather confusion \cite{CGS3,RSsusy}, it is important to comment about this issue regarding
the solutions we have constructed. Concerning the brane located at the fixed point
of the $T$ symmetry, we are missing some elements to make any statement. However
the conclusions are positive for the bulk and the $p$-brane
located at the origin: since locally they correspond to the usual solutions
encountered in supergravity theories, they preserve eight supercharges. If the
second brane breaks part of these supercharges, it would be interesting
to study the transmission of this breaking to the first brane.

In conclusion, we hope that our construction has shed light on the geometrical
origin of the gravity trapping scenario proposed by Randall and Sundrum. It provides
insight on how to extend it to higher-codimensional brane worlds. We have studied
an explicit realization in supergravity models exhibiting finite lower dimensional
Planck mass on the brane despite the non-compact transverse space. Our solution
is invariant under a $T$ symmetry that exchanges the short distances with the
large ones in the transverse space. Two singularities occur that are interpreted as
a $p$-brane at the origin and, at a finite distance, the jump brane,
a ($D-2$)-brane with $\dperp-1$ compact dimensions. The bulk, as well as the
brane at the origin where the
standard model can propagate, preserves half of the sixteen supercharges.
The warp factor is maximum on the jump brane that can be taken as a
{\it Planck brane}, {\it i.e.} a brane where the energy scale would be of the order
of the Planck scale ($\sim 10^{19}$ GeV). In this case, the natural energy scale
on the brane at the origin would be suppressed by a factor $A(r_\circ)/A(0)$ which
can lead to the electroweak scale depending on the location of the $T$ self-dual
point $r_\circ$.
A dynamical description of the jump brane should help to address some interesting
questions like the stabilization of its position, the supersymmetry breaking
transmission to the first brane and the determination of the KK spectrum associated
to the 4D graviton which would allow to compute the deviations to the Newton's law.

\section*{Acknowledgements}

I would like to thank my two former collaborators, J.M. Cline
and G. Servant, for stimulating discussions.
This work was supported in part by the Director, Office
of Energy Research, Office of High Energy and Nuclear Physics, Division of
High Energy Physics of the U.S. Department of Energy under Contract
DE-AC03-76SF00098.


\end{document}